# Dispersive readout with two orthogonal modes of a dielectric cavity

A.M. Kozodaev[1,2,3], I.S. Cojocaru[1,3,4], S.M. Drofa[1,2,3], P.G. Vilyuzhanina[1,2,3], A. Chernyavskiy[1,2,3],

V.G. Vins[6], A.N. Smolyaninov[4], S.Ya. Kilin[5,7], S.V. Bolshedvorskii[3,4], V.V. Soshenko[3,4] and

A.V. Akimov[1,3,4]

[1]*Russian Quantum Center, Bolshoy Boulevard 30, building 1, Moscow, 143025, Russia*
[2]*Moscow Institute of Physics and Technology, 9 Institutskiy per., Dolgoprudny, Moscow Region, 141701, Russia*
[3]*P.N. Lebedev Institute RAS, Leninsky Prospekt 53, Moscow, 119991, Russia*
[4]*LLC Sensor Spin Technologies, 121205 Nobel St. 9, Moscow, Russia*
[5]*National Research Nuclear University "MEPhI", 31, Kashirskoe Highway, Moscow, 115409 Russia*
[6]*LLC Velman, 1/3 st. Zelenaya Gorka, Novosibirsk,630060, Russia*
[7]*B.I. Stepanov Institute of Physics NASB, 68, Nezavisimosty Ave, Minsk, 220072 Belarus*

email: a.akimov@rqc.ru

Nitrogen-vacancy color centers in diamond have proven themselves as a good, sensitive element for the measurement of magnetic fields. While the mainstream of magnetometers based on NV centers uses so-called optically detected magnetic resonance, there has recently been a suggestion to use dispersive readout of a dielectric cavity to enhance the sensitivity of magnetometers. Here, we demonstrate that the dispersive readout approach can be significantly improved if a two-channel scheme is considered.

## I. INTRODUCTION

Today, many scientific groups are working with quantum sensors based on nitrogen-vacancy (NV) color centers in diamond. Numerous theoretical and experimental studies have already focused on reading the states of NV centers optically, by analyzing the luminescence signal of an optically detectable resonance [1–3]. Recently, however, new work has begun to emerge on reading the states of NV centers by measuring the frequency-dispersive shift in a resonator, which can improve sensitivity to magnetic fields [4,5].

When making measurements, it is sometimes advantageous to measure not the direct response of the quantum system, but the response of the resonator in which the quantum system is placed [6]. In this approach, the resonator is exposed to a microwave field, and its transmission or reflection is monitored depending on the state of the quantum system; thus, the transmission depends on the effective frequency of the resonator. This technique is known as dispersive readout.

One of the first theoretical papers on this topic was [7], in which the authors introduced a general input-output theory for quantum dissipative systems, allowing the output of the resonator to be connected to its input through the internal dynamics of the system. Later, in [8], an expression for the resonator transmission coefficient was derived beyond the rotating wave approximation, and the theory of linear response was described. Paper [9] discusses in detail the theory of linear response in a non-Markovian environment. Most analytical calculations are based on the Jaynes-Cummings model, which allows system parameters to be conveniently defined and a linear response function to be obtained without cumbersome calculations. However, it is important to note that this model does not account for collective effects and assumes the presence of only a single (in the simplest case, two-level) quantum system inside the resonator. Thus, there are now many analytical approaches available for describing dispersive readout.

The technique of dispersive readout is used in a number of papers to create quantum magnetic field sensors based on both NV centers and other compounds. As an example of such a sensor, we can cite the work in [10], where the authors used Cr3+ defects in a sapphire crystal (Al2O3) and developed a device with a sensitivity of $9.7 \, \text{pT}/\sqrt{\text{Hz}}$.

It was shown in [5] that the dispersive readout of NV centers in a resonator made it possible to achieve sensitivity beyond the shot-noise limit of optical readout. Since the readout system does not directly interact with the quantum system, it does not destroy the quantum state. According to the authors, this gives the dispersive readout method an advantage over the optical one.

In [4], the possibility of creating a magnetic sensor based on NV centers was investigated, which, according to the authors, would allow the development of a device with a sensitivity of $3 \, \text{pT}/\sqrt{\text{Hz}}$. An important feature of the sensors described above is the significant reduction in noise and readout errors due to the absence of optical interaction with the quantum system.

In this paper, we propose a new readout method using a resonator with two orthogonal modes [11]. During dispersive readout, the phase of the reflected signal is measured. From this signal, the useful component carrying information about the magnetic resonance of the quantum system must be extracted. The use of a resonator with two geometrically orthogonal modes will make it easier to isolate the useful signal and reduce the effect of shot noise on sensitivity by at least a factor of four

## II. THE PROBLEM

The NV center is a defect in the diamond crystal lattice, consisting of a nitrogen atom substituting for a carbon atom and a vacancy at a neighboring lattice site (missing carbon atom) [12]. The energy levels of the NV center at room temperature are shown in Figure 1a. The ground state $^3A_2$ and the excited state $^3E$ are triplets with total spin $S=1$. There are two singlet levels located between the ground and excited states. Decay to the singlet sublevels is possible only from the electronic magnetic sublevels $m_S = \pm 1$. The transition from the singlet state to the ground state $^3A_2$ occurs non-radiatively, primarily into the $m_S = 0$ sublevel. This enables optical pumping into this sublevel.

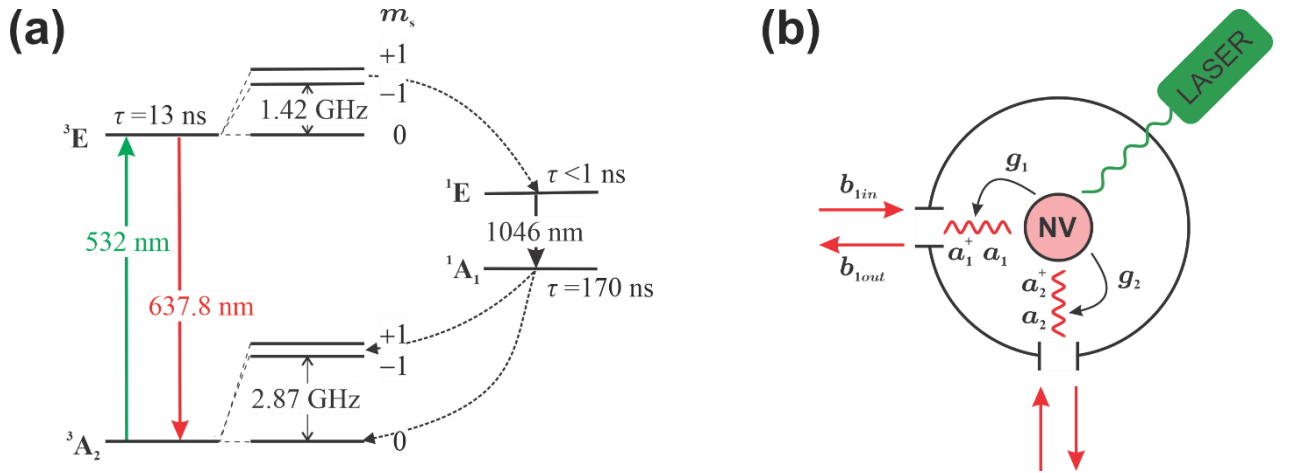

Figure 1. a) Energy level diagram of the NV center at room temperature. The green arrow indicates optical pumping; the red arrow shows fluorescence; the black arrow represents photon emission in the IR range; and the black wavy arrows indicate non-radiative transitions. (b) Resonator circuit with orthogonal modes; a vacuum state is applied to port b_(2,in).

The schematic diagram of the device in question is shown in Figure 1b. A diamond sample containing NV centers is placed in a dielectric resonator [11,13], which creates a uniform microwave field across the NV center ensemble. To polarize the ensemble into a known spin state, continuous laser radiation with a wavelength of 500–550 nm is applied. The resonant frequency of the dielectric resonator is close to the ground-state transition frequency of the NV center $|0\rangle \to |1\rangle$. It is assumed that a small (10–20 G) magnetic field is applied to the system, which makes it possible to isolate the components $|+1\rangle$ and $|-1\rangle$ NV of the ground state of the NV center. An excitation microwave field is applied to the input of resonator 1, while input 2 can be used to detect the magnetic field signal corresponding to the mode orthogonal to the excitation mode. To construct an analytical description of the system, we first consider a resonator with a single mode.

### III. THE HAMILTONIAN

To construct an analytical description of the system, we first write the Hamiltonian using the Jaynes–Cummings model. This model allows for convenient specification of system parameters and facilitates obtaining a linear response function without cumbersome calculations. To describe the quantum system, we consider a setup consisting of an atom $H_{sys}$, a resonator $H_{cav}$, and an external radiation $H_{ex}$:

$$H = H_{sys} + H_{cav} + H_{ex} + H_{I1} + H_{I2}, \tag{1}$$

where $H_{I1}$ is the term responsible for the interaction of the resonator with the external field, and $H_{I2}$ is the term responsible for the interaction of the resonator with the quantum system:

$$\begin{aligned} H_{I1} &= \sum_k \gamma_k \left( b_k^+ a_k + b_k a_k^+ \right) \\ H_{I2} &= gZ\left(a + a^+\right) \end{aligned} \tag{2}$$

In the simplest case, to describe an atom, we use the approximation of a two-level system with the Hamiltonian, $H_{sys}$ which can be written as:

$$H_{sys} = \frac{1}{2}\omega_{sys}\sigma_z, \tag{3}$$

where $\omega_{sys}$ is the transition frequency between the magnetic sublevels $m_s = 0$ and $m_s = -1$ of the ground state of the NV center. The resonator will be considered a single-mode $H_{cav} = \omega_{cav} a^+ a$, where $\omega_{cav}$ is the frequency of the resonator mode, and $a^+$, $a$ are the photon generation/annihilation operators in the resonator mode. The external field is given by the expression:

$$H_{ex} = \sum_k \omega_k b_k^+ b_k, \tag{4}$$

where $\omega_k$ is the frequency of the $k$ mode of the external field, and $b_k^+, b_k$ are the photon creation/annihilation operators in the $k$ mode of the external field. The resulting Hamiltonian has the form:

$$H = \frac{1}{2}\omega_{sys}\sigma_z + \omega_{cav}a^+a + \sum_k \omega_k b_k^+ b_k + \sum_k \gamma_k \left(b_k^+ a_k + b_k a_k^+\right) + gZ\left(a + a^+\right). \tag{5}$$

## IV. THE LANGEVIN EQUATION

From the Hamiltonian obtained above, linearized Langevin quantum equations can be derived, which determine the dynamics of the system. We will work in Heisenberg's view:

$$\begin{aligned}\dot{a} &= -i\omega_{cav}a - igZ - \frac{\varkappa}{2}a - \sum_v b_{in,v}\sqrt{\varkappa_v} \\ \varkappa &= 2\pi \sum_k \gamma_k^2 (\omega_k - \omega_{cav}) \\ \varkappa_1 + \varkappa_2 &= \varkappa, \quad b_{out,v} - b_{in,v} = \sqrt{\varkappa_v}\, a\end{aligned} \tag{6}$$

The value $\varkappa$ is a loss coefficient, which in the case of a cavity, defines the quality factor $Q = \omega_{cav}/\varkappa$, $\varkappa_1$, $\varkappa_2$ which represents the input and output losses of the resonator, respectively.

Since we are not interested in quantum fluctuations of the resonator field, we consider equation (6) in its classical limit, as the equation of motion for the average values of $a \equiv \langle a \rangle_t$ and $Z \equiv \langle Z \rangle_t$. To further describe the system, we express $Z$ a using the linear response function, using the results of [8]. The sensitivity or linear response function of a system describes how and with what delay a quantum system reacts to a weak external influence; in addition, it depends only on the properties of the system, and not on the resonator. According to [8], for a multilevel quantum system, it has the form:

$$\chi(\omega) = \sum_{m,n} \frac{(p_m - p_n)|Z_{mn}|^2}{\omega + E_m - E_n + i\frac{\gamma_{mn}}{2}}, \tag{7}$$

where $\gamma$ is the rate of coherence loss, and $p_n$ is the population of the energy levels of the system. In equation (6), we replace $a$ and $a^+$ with the classical amplitudes $a$ and $a^*$ and perform the Fourier transform. Note that to obtain a strong signal, the resonator must have a high Q-factor

$Q = \dfrac{\omega_{cav}}{\varkappa} \gg 1$, must be close to the resonance $|\omega - \omega_{cav}| \ll \omega_{cav}$, and the dispersive shift $\delta\omega = g^2 Re\,\chi(\omega)$ must be less than the resonator frequency. In total, we get the inequalities:

$$\varkappa,\ |\omega - \omega_{cav}|,\ g^2 Re\,\chi(\omega) \ll \omega_{cav}. \tag{8}$$

Then the Langevin equation takes the form:

$$i(\omega - \omega_{cav})a_\omega - ig^2 a_\omega \chi(\omega) - \dfrac{\varkappa}{2} a_\omega = \sum_v b_{in,\,v} \sqrt{\varkappa_v}. \tag{9}$$

Together with the equations for $b_{in}$ and $b_{out}$ from equation (9), we obtain the transmission and reflection coefficients, respectively:

$$t = \dfrac{b_{out,2}}{b_{in,1}} = \dfrac{i\sqrt{\varkappa_1 \varkappa_2}}{\omega_{cav} - \omega + g^2\chi(\omega) - i\dfrac{\varkappa}{2}}, \tag{10}$$

$$r = \dfrac{b_{out,1}}{b_{in,1}} = 1 + \dfrac{i\,\varkappa_1}{\omega_{cav} - \omega + g^2\chi(\omega) - i\dfrac{\varkappa}{2}}. \tag{11}$$

To calculate the populations, we will numerically solve the Lindblad equation:

$$\dot\rho = -i[H,\rho] + \sum_j L_j \rho L_j^+ - \dfrac{1}{2}\{L_j^+ L_j, \rho\}. \tag{12}$$

As the Hamiltonian, we take:

$$H = -\dfrac{1}{2}\omega_{sys}\sigma_z - \Omega_R \sigma_x \cos\omega t, \tag{13}$$

but to speed up the calculations, we switch to the rotating wave approximation RWA:

$$H_{RWA} = \dfrac{\omega - \omega_{sys}}{2}\sigma_z - \dfrac{\Omega_R}{2}\sigma_x. \tag{14}$$

Dissipators $L_j$ are responsible for pumping and relaxing the levels of the system. They are determined by the characteristic times of optical pumping, relaxation from the excited state, thermal excitation, and the decoherence time. For each detuning value, we will calculate the stationary population value (Figure 2a). Then, for each setting value and its corresponding population, we will determine the value of the transmission coefficient (Figure 2b).

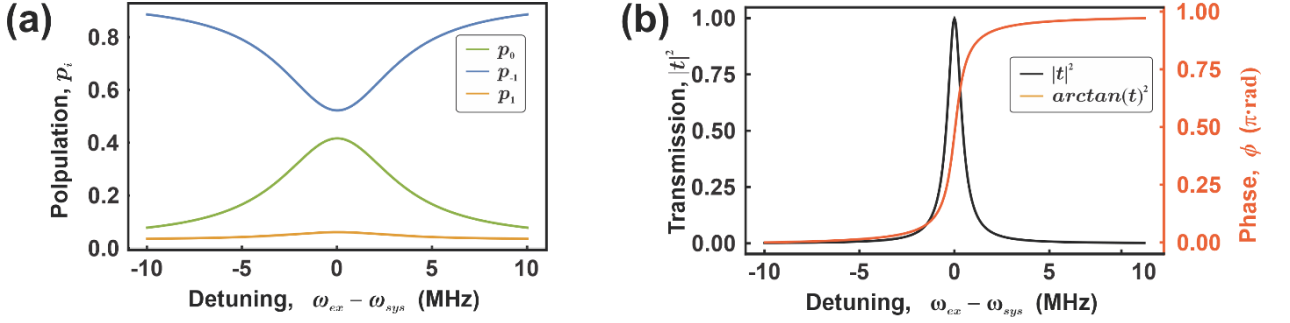

*Figure 2 a) The dependence of the population of the ground state levels on the detuning in the numerical solution of the Lindblad equation; b) The dependence of the transmission coefficient and phase on detuning.*

So far, we have considered a single-mode resonator. As a method to improve the frequency readout of the system in the resonator, we propose using a resonator with two geometrically orthogonal modes. The idea behind using orthogonal modes is to measure the transmission coefficient of the resonator from port 1 to port 2. Provided that the external fields in the two ports of the resonator are each connected to their own mode, the transmission should be zero in the case when the field frequency is strongly tuned from the resonance frequency of the two-level system. When the frequency of the field and the two-level system coincide, scattering will appear on the two-level system and modes 1 ($a_1$) into mode 2 ($a_2$), and as a result, a signal will appear at output 2. To determine the type of interaction of spin-1/2 with the orthogonal mode, we will act on the initial interaction Hamiltonian by the rotation operator $U$:

$$U = \exp\left(-i\sigma_z \frac{\pi}{4}\right). \tag{15}$$

Then we get:

$$U\left(\sigma_+ a + \sigma_- a^+\right)U^+ = i\left(\sigma_+ a - \sigma_- a^+\right). \tag{16}$$

In this case, the full interaction Hamiltonian will have the form:

$$H_{int} = g_1\left(\sigma_+ a_1 + \sigma_- a_1^+\right) + ig_2\left(\sigma_+ a_2 - \sigma_- a_2^+\right). \tag{17}$$

The Langevin equalizations corresponding to modes $a_1$ and $a_2$ are equivalent to (9):

$$i(\omega_{ex} - \omega_1)a_1 - ig_1\chi(g_1 a_1 + ig_2 a_2) - \frac{\varkappa_1}{2}a_1 = b_{1,in}\sqrt{\varkappa_1}, \tag{18}$$

$$i(\omega_{ex} - \omega_2)a_2 - g_2\chi(g_1 a_1 + ig_2 a_2) - \frac{\varkappa_2}{2}a_1 = b_{2,in}\sqrt{\varkappa_2}, \tag{19}$$

$$b_{1,out} - b_{1,in} = a_1\sqrt{\varkappa_1}, \quad b_{2,out} = a_2\sqrt{\varkappa_2}, \quad b_{2,in} = 0. \tag{20}$$

In this system, we are interested in $S_{21} = b_{2,out}/b_{1,in}$, the expression for which has the form:

$$S_{21} = \frac{b_{2,out}}{b_{1,in}} = \frac{g_1 g_2 \chi \sqrt{\varkappa_1 \varkappa_2}}{\left(i(\omega_{ex}-\omega_1) - ig_1^2\chi - \frac{\varkappa_1}{2}\right)\left(i(\omega_{ex}-\omega_2) - ig_2^2\chi - \frac{\varkappa_2}{2}\right) + g_1^2 g_2^2 \chi^2}. \tag{21}$$

## V. THE SENSITIVITY

Above, we considered two methods for determining the frequency of a quantum system in a resonator, based on measuring the phase change of a signal transmitted through the resonator, specifically, the method using a single mode and the method using two orthogonal modes. In this section, we compare both methods to determine which one introduces less error in measuring the frequency of the quantum system.

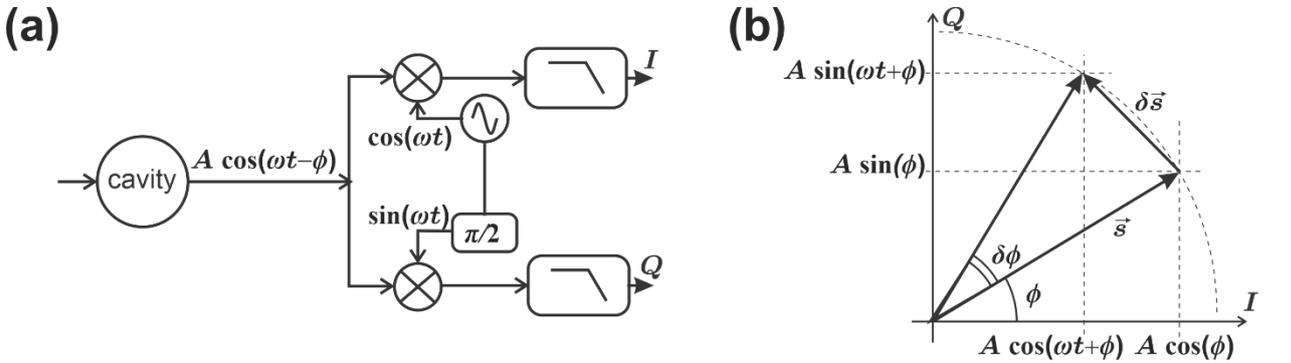

*Figure 3 (a) The operation scheme of the IQ demodulator. (b) Schematic representation of the phase change on the graph in space (I, Q)*

Let an external field be incident on the resonator, acquiring a phase shift $\varphi$ after passing through the resonator $A\cos(\omega t - \varphi)$ Figure 3. We multiply the signal by the sin and cos of the carrier frequency, then apply a low-pass filter and get two signals $I$ and $Q$:

$$I = A\sin\varphi, \quad Q = A\cos\varphi, \tag{22}$$

$$\text{RF} = A\cos(\omega t - \varphi) = A\cos\omega t \cos\varphi + \sin\omega t \sin\varphi = I\cos\omega t + Q\sin\omega t, \tag{23}$$

$$\text{RF}\cos\omega t = I\cos^2\omega t + Q\sin\omega t \cos\omega t = \frac{I}{2}(1+\cos 2\omega t) + \frac{Q}{2}\cos 2\omega t \rightarrow \text{LPF} \rightarrow I, \tag{24}$$

$$\text{RF}\sin\omega t = Q\cos^2\omega t + I\sin\omega t \cos\omega t = \frac{Q}{2}(1-\cos 2\omega t) + \frac{I}{2}\cos 2\omega t \rightarrow \text{LPF} \rightarrow Q. \tag{25}$$

At the output of the demodulator, we get a common-mode signal $I$ and a quadrature $Q$, which together give the final signal $S = I + iQ = Ae^{i\varphi}$. To take into account the presence of phase noise, we introduce an additional phase shift $\delta\varphi \ll 1$, which corresponds to the displacement signal (phase noise signal) $\delta S = S\delta\varphi$ (Figure 3(b)). Thus, the phase measurement error caused by the phase noise of the demodulator has the form $\delta\varphi = \delta S/S$.

Errors in determining the in-phase and quadrature signals $\delta I$ and $\delta Q$ occur due to the presence of noise. We will determine $\delta I$ and $\delta Q$ based on the characteristics of the system and the demodulator. The quadrature IQ demodulator is characterized by the Noise Figure (NF) parameter. For the existing IQ demodulator for the input field with a frequency of 2800 MHz, the NF parameter is within $NF \in (12,15)$ dB. The NF value determines the quotient of the signal-to-noise ratio (SNR) at the input to the modulator and the output:

$$NF = 10\log_{10}\frac{SNR_{in}}{SNR_{out}}. \tag{26}$$

The signal-to-noise ratio is defined by the ratio of the current signal power $S$ with payload $R$ to the current noise power $\delta S$:

$$SNR_{out} = \frac{S^2 R}{\delta S^2 R} = \frac{S^2}{\delta S^2} = \frac{1}{\delta\varphi^2}. \tag{27}$$

When determining $SNR_{in}$, we will consider only shot noise, as it dominates over other noise sources in this system, as shown in [10]. For an ideal detector, shot noise sets the limit of accuracy. The standard quantum limit, according to [14] is defined as $\delta\varphi = 1/\sqrt{N}$. Then $SNR_{in} = N$, where $N$ is the number of photons that hit the detector during the observation period $\tau = 1\,\text{s}$. Then $N$ is:

$$P_{out} = |t|^2 P_{in} = \frac{N\hbar\omega_c}{\tau} \rightarrow N = \frac{|t|^2 P_{in} \tau}{\hbar\omega_c}.$$

Thus, the error of measuring the phase of the signal $\delta\varphi$ will be:

$$\delta\varphi = \frac{10^{NF/20}}{\sqrt{N}}.$$

We will define the sensitivity of the shot noise $\eta_{SN} = \delta B = \delta\omega/\gamma_e$ as being proportional to the error between the measured frequency of the quantum system $\omega_m$ and the actual frequency $\omega_{real}$:

$$\omega_m = \omega_{real} + (\varphi_m - \varphi_{real})\frac{d\omega_{sys}}{d\phi}, \qquad (28)$$

$$\delta\omega = \delta\varphi \left(\frac{d\phi}{d\omega_{sys}}\right)^{-1} = 10^{\frac{NF}{20}}\sqrt{\frac{\hbar\omega_c}{|t|^2 P_{in}\tau}}\left(\frac{d\phi}{d\omega_{sys}}\right)^{-1} \qquad (29)$$

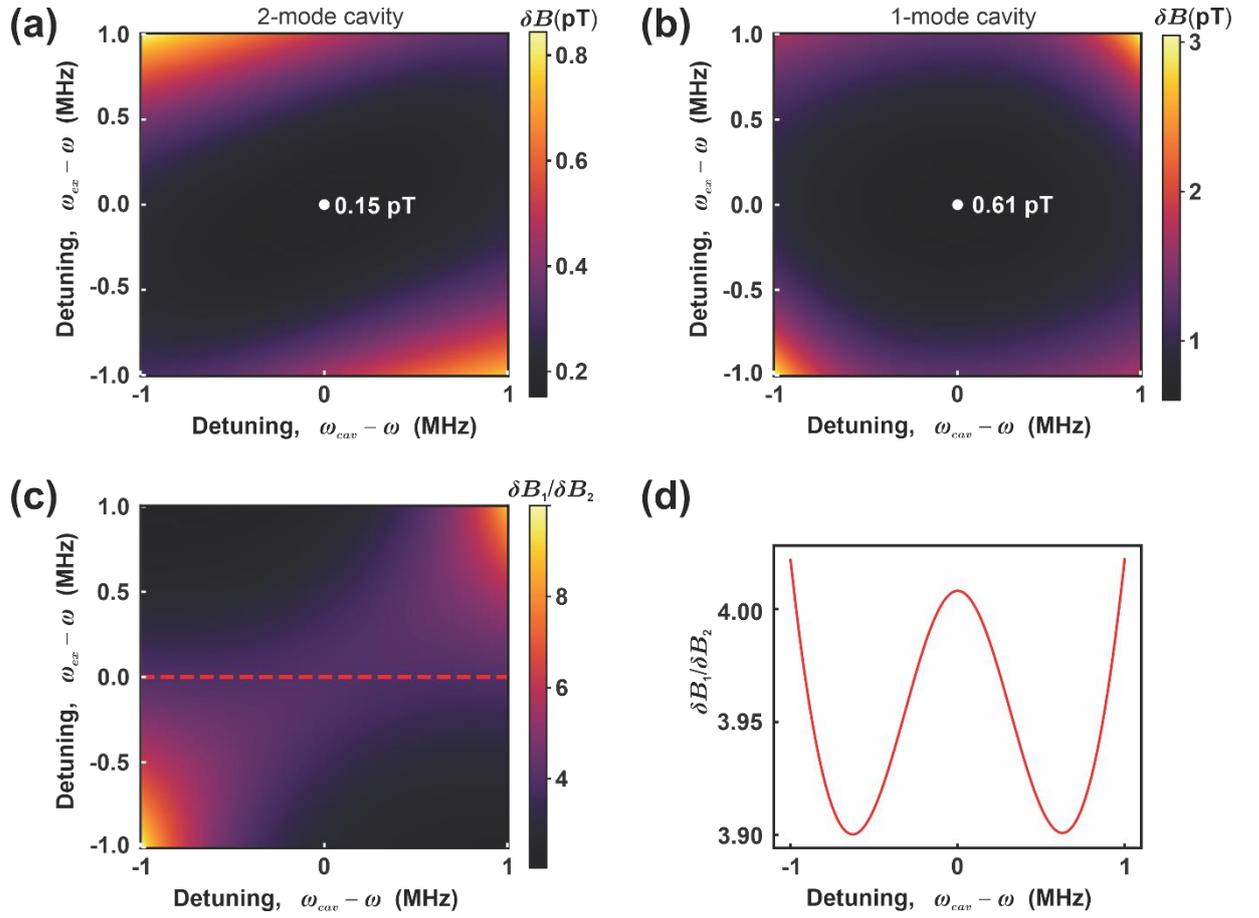

*Figure 4 a) Estimation of sensitivity by photon shot noise for the 1-mode case; b) for the 2-mode case; c) sensitivity ratio for the 1- and 2-mode cases; d) sensitivity ratio for the 1- and 2- mode cases when resonating microwaves with NV.*

The obtained estimate based on shot noise $\eta_{SN}(\omega_{cav}-\omega_{sys}, \omega_{ex}-\omega_{sys})$ is represented in the heat maps of Figure 4(a,b). Simulation parameters: Rabi frequency $0.2\times 2\pi$ MHz, resonator amplitude $Q=5000$, microwave radiation power 40 mW. A sample with dimensions of $1\times 1\times 0.5\,\text{mm}^3$ and a total number of NV centers equal to $9\times 10^{13}$, was selected for modeling. The sensitivity minima occur at the point of full resonance and are equal to 0.61 pT and 0.15 pT for the 1- and 2-mode cases, respectively. To compare the obtained sensitivity values, we construct the ratio $\eta_{SN1}/\eta_{SN1}$ in Figure 4(c) and take a slice at the point of minimum sensitivity, where the external field frequency coincides with the NV resonance (indicated by the red line in the figure). Figure 4(d) shows that at the optimal resonant point for measurement, the magnetic field sensitivity of a resonator with two orthogonal modes is more than four times greater than that of a resonator with a single mode.

## VI. CONCLUSIONS

The paper considers a method for the dispersive readout of a magnetic field signal from a dielectric resonator + nitrogen vacancy centers in diamond. The method is based on detecting changes in the resonator state, which vary with the external magnetic field due to the coupling of the resonator with nitrogen vacancy centers. It is shown that reading the magnetic field signal from the response of the NV-centers–resonator system through a mode orthogonal to the excitation mode provides a fourfold improvement in signal strength compared to direct dispersive readout. This method enables a sensitivity of 0.15 pT using realistic diamond parameters.

## VII. ACKNOWLEDGMENTS

This study was supported by the Ministry of Science and Higher Education of the Russian Federation (agreement/grant No. 075-15-2024-556).